%% file: INTERSPEECH2021.tex
\documentclass[a4paper]{article}

\usepackage{INTERSPEECH2021,balance,url}

\title{Speaker anonymisation using the McAdams coefficient}
\name{Jose Patino$^1$, Natalia Tomashenko$^2$, Massimiliano Todisco$^1$, Andreas Nautsch$^1$ and Nicholas Evans$^1$}
\address{$^1$EURECOM, Sophia Antipolis, France\\
         $^2$LIA, Avignon Universit\'e, France}
\email{$^1$firstname.lastname@eurecom.fr, $^2$firstname.lastname@univ-avignon.fr}

\begin{document}

\maketitle
\begin{abstract}
\input{sections/0_abstract}

\end{abstract}


\section{Introduction}
\label{sec:intro}
\input{sections/1_introduction}

\section{Previous work}
\label{sec:previous_work}
\input{sections/2_previous_work}

\section{Anonymisation using \\the McAdams coefficient}
\label{sec:pseudonymisation}
\input{sections/3_pseudonymisation}

\section{Experiments}
\label{sec:experimental_setup}

\input{sections/4_experimental_setup}

\subsection{Results}
\label{sec:results}

\input{sections/5_results}

\section{Conclusions}
\label{sec:conclusions}
\input{sections/6_conclusions}

\section{Acknowledgements}

This work is partly funded by the VoicePersonae project which is supported by the French Agence Nationale de la Recherche (ANR) and the Japan Science and Technology Agency (JST).  It is also linked to the VoicePrivacy initiative and the Harpocrates project also funded by the ANR.

\balance
\bibliographystyle{IEEEtran}

\bibliography{INTERSPEECH2021}

\end{document}

%% file: sections/0_abstract.tex
Anonymisation has the goal of manipulating speech signals in order to degrade the reliability of automatic approaches to speaker recognition, while preserving other aspects of speech, such as those relating to intelligibility and naturalness.  
This paper reports an approach to anonymisation that, unlike other current approaches, requires no training data, is based upon well-known signal processing techniques and is both efficient and effective.  The proposed solution uses the McAdams coefficient to transform the spectral envelope of speech signals.  Results derived using common VoicePrivacy 2020 databases and protocols show that random, optimised transformations can outperform competing solutions in terms of anonymisation while causing only modest, additional degradations to intelligibility, even in the case of a semi-informed privacy adversary.

%% file: sections/1_introduction.tex
Recent years have seen an increase in privacy legislation.  Much of it covers what is referred to as \emph{personally identifiable information} (PII), e.g.\ biometric data such as speech~\cite{nautsch2019gdpr,nautsch2019preserving}.  The community-led VoicePrivacy initiative~\cite{tomashenko2020introducing} aims to foster progress in the development of anonymisation techniques that can be employed to conceal PII contained within speech signals.  

The VoicePrivacy 2020 evaluation plan~\cite[Sec.~3.2] {tomashenko2020voiceprivacy} specifies four requirements for successful anonymisation solutions.  They should: (i)~produce a speech waveform; (ii)~suppress speaker-specific information as much as possible; (iii)~preserve intelligibility and naturalness; (iv)~protect voice distinctiveness. Anonymised speech should degrade the reliability of identification applications such as automatic \emph{speaker} recognition, but should not interfere with the correct functioning of utility applications such as automatic \emph{speech} recognition.  Anonymised speech recordings originating from the same source speaker should have the same voice, while the linking of anonymised voices to source speakers should be prevented.  Given that each speaker should retain a distinctive, but anonymised voice, the task is perhaps referred to more appropriately as pseudonymisation~\cite{noe2020speech}.

Best practice dictates that assessment is applied according to different attack models or scenarios whereby a privacy adversary has knowledge of, or access to, a similarly (but not identically configured) anonymisation system. To support assessment according to the various VoicePrivacy scenarios, anonymisation is required to be applied separately, with different configurations, to two different data subsets, enrolment and test, such that the anonymised voice for each speaker in each set is different.  Further details concerning the attack models are discussed in~\cite[Sec.~3.3]{tomashenko2020voiceprivacy}.

Two baseline anonymisation systems\footnote{\url{https://github.com/Voice-Privacy-Challenge/Voice-Privacy-Challenge-2020}} 
were made available to VoicePrivacy~2020 participants: (i)~a primary baseline based upon state-of-the-art x-vector embeddings and neural waveform techniques~\cite{fangspeaker}; 
(ii)~a secondary baseline, inspired from the McAdams coefficient~\cite{mcadams1984spectral}, consisting of well-known signal processing techniques~\cite{patino2020mcadams}.  The primary baseline is comparatively complex and requires substantial training data and computational resources.  Based upon a simple contraction or expansion of pole locations derived using linear predictive coding (LPC), the secondary baseline requires no training data and is comparatively straightforward and efficient.  While the primary baseline better suppresses speaker specific information~\cite{tomashenko2020introducing} (requirement~ii), the secondary baseline better preserves voice distinctiveness~\cite{noe2020speech} (requirement~iv).  The secondary baseline was made available to provide additional inspiration, to expose a broader potential solution space and to lower the cost of entry to the VoicePrivacy initiative.  The intention was to help attract potential participants with a signal processing background that might lack the background in speaker characterisation required to explore the primary baseline.  The secondary baseline was not optimised in any way (other than through casual listening tests), was applied in a deterministic fashion such that anonymisation is easily reversible and was adopted by others~\cite{gupta2020design,kai2021slt}.

The work reported in this paper aims (i)~to explore more thoroughly the potential of well-known signal processing techniques as a solution to the anonymisation problem, and (ii)~to assess the real need for and benefit of more complex, demanding solutions.  The paper reports our efforts to optimise the original McAdams-based solution and its adaptation to a more stochastic approach which affords better protection from reversibility.  Longer-term, it is hoped that this work might expose opportunities to improve performance by combining the components or techniques used by each baseline solution.

The remainder of this paper is organised as follows. 
Related past work is described in Section~2. The proposed approach to anonymisation is introduced in Section~3.  Experiments are reported in Section~\ref{sec:experimental_setup} while a discussion of findings, conclusions and ideas for future work are presented in Section~\ref{sec:conclusions}.

%% file: sections/2_previous_work.tex
A popular synthetic sound generation techniques in the field of music signal processing is that of additive synthesis~\cite{computermusic}. 
The technique is used to generate timbre through the addition of multiple cosinusoidal oscillations: 

\begin{equation}
    y ( t ) = \sum _ { k = 1 } ^ { K } r _ { k } ( t ) \operatorname { cos } ( 2 \pi (k f _ { 0 }) ^ {\alpha} t + \phi _ { k } )
    \label{eq:mcadams}
\end{equation}

\noindent where $k$ is the harmonic index, $r _ { k } ( t )$ is amplitude,
$\phi _ { k }$ is the phase, $t$ is time and $\alpha$ is the so-called McAdams coefficient~\cite{mcadams1984spectral}.  Equation~\ref{eq:mcadams} resembles the synthesis of a periodic signal with an inverse Fourier series which combines harmonic cosinusoidal oscillations, each with some magnitude and some phase shift.

The McAdams coefficient is used to adjust the frequency of each harmonic.  Adjustments to the distribution of harmonics acts to modify the resulting timbre.  When applied to additive synthesis as in (\ref{eq:mcadams}), the McAdams coefficient transforms harmonics to non-harmonic components which are also referred to as \emph{partials} or overtones.

This technique bears similarities to other, well-known techniques in speech processing~\cite{jin2009speaker, pobar2014online, bahmaninezhad2018convolutional, qian2017voicemask,magarinos2017reversible,srivastava2019privacy}.  One example is vocal tract length normalization (VTLN)~\cite{cohen1995vocal, lee1998frequency} which has been explored previously in anonymisation-related work~\cite{qian2018hidebehind,srivastava2020evaluating}.  Warping functions are applied to the spectral log amplitude of each speech frame to modify not just the envelope, as in our approach described in Section~3, but also the pitch.  It can hence be argued that these approaches offer greater anonymisation strength and that our approach may then be inferior where identification is performed by human listeners.  However, conversion of the pitch may be somewhat redundant if anonymisation is to be performed purely for the purposes of deceiving \emph{automatic} approaches to identification which rarely use estimates of pitch or source/residual signals.

%% file: sections/3_pseudonymisation.tex
\begin{figure}[!t]
    \centering
    \includegraphics[width=\columnwidth]{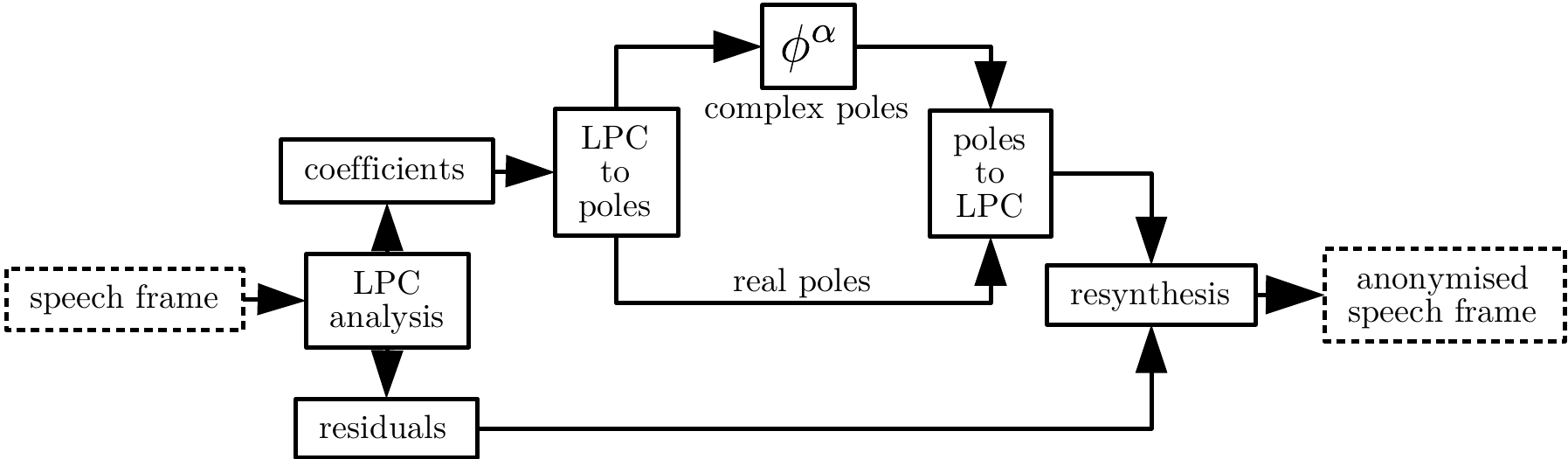}
    \caption{Pipeline of the application of the proposed McAdams coefficient-based approach to anonymisation on a speech frame basis. The angle $\phi$ of poles with a non-zero imaginary part are raised to the power of the McAdams coefficient $\alpha$ to provoke a expansion/contraction in frequency in its associated formant.}
    \label{fig:lpc_processing}
\end{figure}

Our approach to pseudonymisation is explained in this section.  It is based upon the idea of applying a shift to the formant positions in a speech utterance, thereby adjusting the timbre or spectral envelope.  Representations of the latter are often used in the front-end of an automatic speaker verification (ASV) system.  Consequently, manipulations to the formant positions should degrade ASV reliability and hence provide some level of anonymisation.  In our approach, the degree of formant shifting is controlled by the McAdams coefficient~\cite{mcadams1984spectral} as in (\ref{eq:mcadams}).  The principal ideas are outlined first, before we show the resulting system is applied to the anonymisation of speech signals.

\subsection{Application to speech signals}

We use the McAdams coefficient to manipulate the formant positions of speech signals at the frame level.  The system described below is that which we proposed as the secondary baseline for the VoicePrivacy 2020 challenge. The process is illustrated in Figure~\ref{fig:lpc_processing}. First, source-filter analysis is applied to each frame using linear predictive coding (LPC). 
The source or residual is set aside for later resynthesis, whereas filter coefficients are used to derive the set of pole positions.
Real-valued pole positions (with zero-valued imaginary terms) are left unmodified, whereas complex-valued poles (with non-zero imaginary terms) are shifted according to the higher branch in Figure~\ref{fig:lpc_processing}.

The shift, applied via the same transformation in (\ref{eq:mcadams}), operates on the angle between the positive real axis and the vector extending from the origin in the z-plane to the complex pole position.  This angle corresponds to frequency, with the upper half of the unit circle (angle of $\pi$ radians) corresponding to the sampling frequency.  The full set of complex-valued pole positions is shifted according to $\phi^\alpha$ resulting in either clockwise or counter-clockwise shifts
in pole locations.  

\begin{figure}[!t]
    \centering
    \includegraphics[trim={3cm 1.5cm 5cm 4cm},clip,width=1.0\columnwidth]{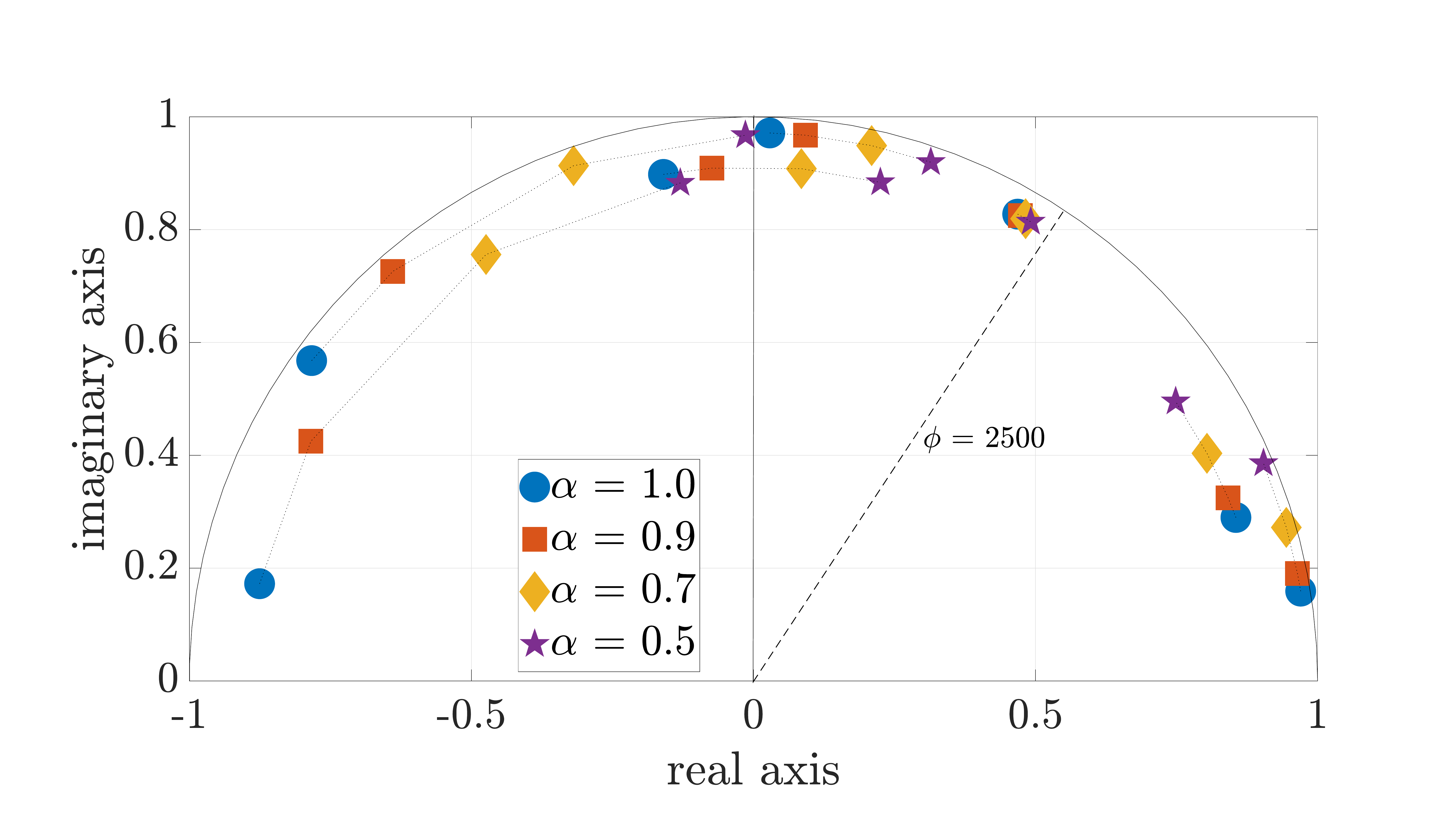}
    \caption{An illustrative example of original pole positions and pole shifts for values of $\alpha\in\{0.9, 0.7, 0.5\}$.  Only the top-half of the pole-zero map is shown.  Complex conjugate poles undergo the same shift, but in the opposite direction. 
    }
    \label{fig:pole-zero_plot}
\end{figure}

This result of this process is illustrated in the z-plane of Figure~\ref{fig:pole-zero_plot} where the blue circles indicate original pole locations and where other points indicate induced shifts.
As shown, the shift is different for each pole in terms of both direction and the scale of the shift. For poles with an angle of $\phi<1$ radian, the shift $\phi^{\alpha}$ is counter-clockwise for values of $\alpha<1$.  While not explored in this work, values of $\alpha>1$ would induce shifts in a clockwise direction for values of $\phi<1$ radian. Poles with an angle of $\phi>1$ radians undergo shifts in the opposite directions. Lastly, the scale of the shift is dependent upon the gap between $\phi$ and $\phi=1$ radian; the greater the gap, the greater the shift.  Example shifts in pole positions are illustrated in Figure~\ref{fig:pole-zero_plot} for values of $\alpha<1$ for values $\alpha\in\{0.9, 0.7, 0.5\}$.

An illustration of the result in terms of spectral envelope is shown in Figure~\ref{fig:mcadams}.  Four spectra are illustrated.  The spectra illustrated by the solid blue line is the original speech spectrum whereas the others show the spectra after anonymisation for values of $\alpha\in\{0.9, 0.7, 0.5\}$.  The nearer a pole to the arc of the unit circle (Figure~\ref{fig:lpc_processing}), the more apparent the corresponding peak in the spectral envelope, with each pole and peak loosely corresponding to a speech formant. Given a sampling rate of 16~kHz (as for VoicePrivacy 2020 data), the value of $\phi=1$ radian corresponds to a frequency of approximately 2.5~kHz.  The shift in pole positions is hence in opposite directions for frequencies either side of this threshold and is either an expansion away from 2.5~kHz or a contraction towards it.

Returning to Figure~\ref{fig:lpc_processing}, the set of new poles, including original real-valued poles and those modified according to the format shifting procedure, are then converted back to LPC coefficients.  The latter are combined with the residual from the original speech signal and then used to resynthesise an anonymised speech frame in the time domain.  The stream of speech frames, each treated according to the procedure described above, is then combined by means of a standard overlap and add (OLA) technique to produce the final anonymised speech signal.

\subsection{Stochastic anonymisation}
\label{subsec:stochasticity}

The original VoicePrivacy baseline used an arbitrary, fixed McAdams coefficient of $\alpha=0.8$.  Clearly, there is scope for optimisation since different values further away from $\alpha=1$ will potentially induce greater shifts in the pole positions and should hence provide greater levels of anonymisation (higher EERs).  However, better anonymisation will likely come at the expense of speech distortion, with the latter likely also causing degradation to intelligibility and naturalness.  Accordingly, we report new work which aims to explore the trade off between these two competing objectives.  We restrict the study to values of $\alpha<1$.  Casual listening tests showed that the use of values of $\alpha>1$ resulted in less acceptable degradations to intelligibility.

Next, rather than using a fixed, deterministic value of $\alpha$, which implies that anonymisation is reversible, we sought to confirm that different values of $\alpha$ can produce different pseudo-voices for the same speaker.  
The goal here is to support experimentation with different attack models and scenarios by ensuring that the anonymisation of enrolment and test data results in \emph{different}, or sufficiently dissimilar pseudo-voices (see Section~1 and~\cite[Sec.~3.2]{tomashenko2020voiceprivacy}).  The use of a fixed $\alpha$ in the original baseline meant that the anonymisation of enrolment and test utterances previously resulted in the \emph{same}, or similar pseudo-voices.

The result is a stochastic approach to anonymisation whereby the McAdams coefficient is sampled from within a range of the uniform distribution, i.e.\ $\alpha \in \text{U}(\alpha_{\text{min}}, \alpha_{\text{max}})$.  The application of anonymisation to enrolment and test data with different, random McAdams coefficients drawn from different ranges of the same uniform distribution should result in different pseudo-voices for each speaker's enrolment and test utterances, while also providing some protection from reversibility.  We note that anonymisation is applied on a speaker-dependent basis, with each speaker having a randomly chosen McAdams coefficient.  A privacy adversary would then need to know the exact McAdams coefficient used to anonymise the speech of any particularly speaker in order to revert the transform.

%% file: sections/4_experimental_setup.tex
This section describes the VoicePrivacy 2020 database used in this work, the metrics used for assessment, the different attack models or scenarios and our results.

\subsection{Data and metrics}

We used the VoicePrivacy 2020 database and protocols described in~\cite{tomashenko2020introducing,tomashenko2020voiceprivacy}.  However, in contrast to the primary baseline~\cite{fangspeaker,srivastava2020design}, our anonymisation system has no requirement for training data.  Consequently, our solution makes no use of the training partition. We report results for evaluation data only, which is drawn from the LibriSpeech~\cite{panayotov2015librispeech} and VCTK~\cite{veaux2016vctk} source datasets.  While we have observed differences in performance for each dataset, as well as gender dependencies, for reasons of space limitations we report only average performance here.\footnote{Full results available at \url{https://github.com/josepatino/Voice-Privacy-Challenge-2020/blob/master/results/}}  Database, gender dependencies and other potential biases, obviously warrant investigation in further work. Assessment is performed using the standard VoicePrivacy 2020 ASV and ASR systems~\cite{tomashenko2020introducing} both trained using the train-clean-360 partition of the LibriSpeech database.

Degradation to intelligibility is measured through the ASR word error rate (WER) whereas anonymisation performance is measured through the ASV equal error rate (EER).  Also reported here are estimates of privacy using the zero evidence biometric recognition assessment (ZEBRA) framework~\cite{nautsch2020privacy}, a recently proposed, adversary-agnostic, metric inspired by the work of the forensic sciences community.
\begin{figure}[t]
    \centering
    \includegraphics[width=\columnwidth]{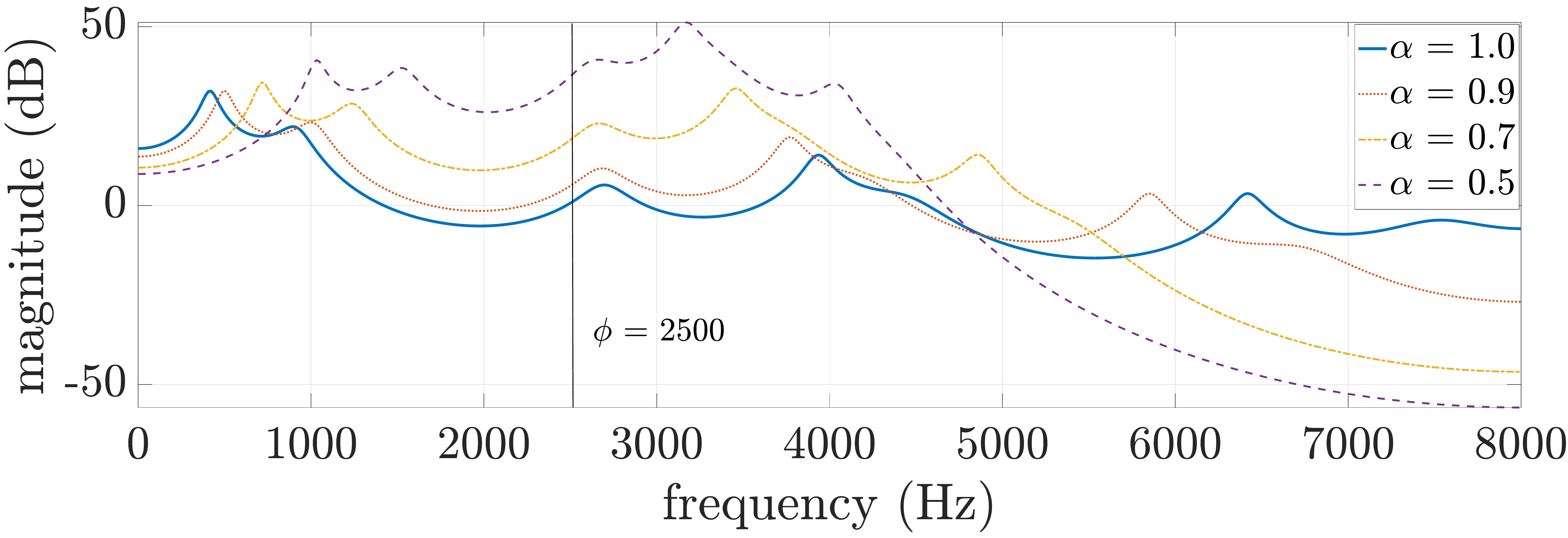}
    \caption{Spectral envelopes for an original speech frame and pole positions, and for anonymised versions for $\alpha\in\{0.9, 0.7, 0.5\}$.  Anonymisation results in an expansion or contraction of the spectrum either side of $\phi=1$~radiant or 2.5~kHz.
    }
    \label{fig:mcadams}
\end{figure}
\subsection{Scenarios}
\label{subsec:scenarios}

In adopting the terminology of~\cite{srivastava2020evaluating}, we investigated anonymisation performance according to two different attack models or scenarios.  Both involve a privacy adversary who seeks to determine whether two utterances belong, or not, to the same speaker.  In both cases there is a test utterance which is anonymised, and an enrolment utterance which is either an original utterance, or a similarly-anonymised utterance.  

The first scenario assumes an~\emph{ignorant} privacy adversary who has no knowledge of the test utterance being anonymised.  In this case the adversary attempts to determine the match between an {\bfseries o}riginal enrolment utterance and an {\bfseries a}nonymised test utterance.  This is referred to as the {\bfseries o-a} scenario.  The second scenario assumes a \emph{semi-informed} privacy adversary who, while knowing that the test utterance has been anonymised and having access to the systems used, does not know the configuration.  In this case, the adversary will likely determine more reliably the match between the two utterances by anonymising the enrolment utterance such that the comparison is now between two similarly anonymised utterances, even though each is anonymised with differently configured systems. It is referred to as the {\bfseries a-a} scenario. Like~\cite{srivastava2020evaluating}, we do not consider the unlikely scenario of an \emph{informed} adversary that has knowledge of both the algorithm and the configuration.

We also report ASV and ASR results derived from systems that are retrained on either original or anonymised data.  These experiments aim to determine robustness in the case that the adversary is able to harness knowledge of the anonymisation system to break the protection it affords.  Similarly, we seek to determine the improvements one can obtain in intelligibility, if the ASR system is adapted to anonymised data.  See~\cite{tomashenko2020voiceprivacy} for further details.

%% file: sections/5_results.tex
\begin{figure}[!t]
    \centering
    \includegraphics[width=\columnwidth]{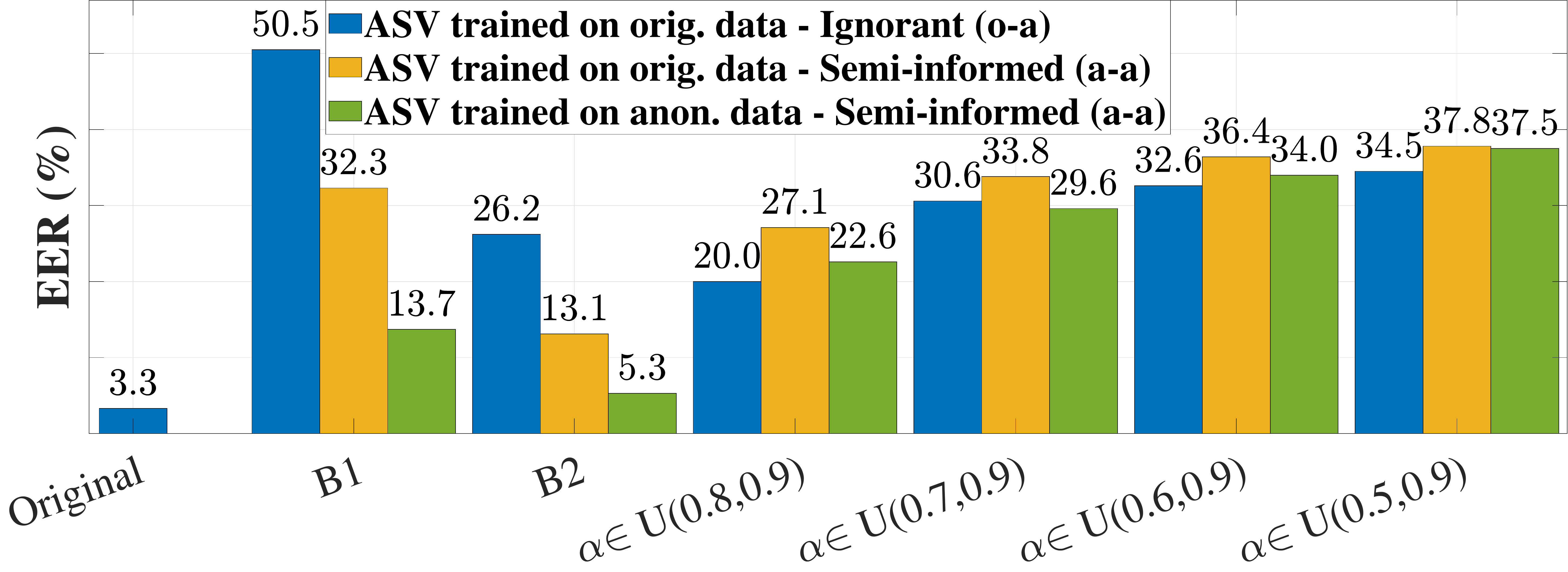}
    \caption{ASV performance in terms of the equal error rate (EER, \%) for various systems and for the VoicePrivacy 2020 test set. B1 is the primary baseline as in~\cite{tomashenko2020introducing,fangspeaker,srivastava2020design}, while B2 is the secondary baseline as in~\cite{patino2020mcadams} with a fixed $\alpha=0.8$.  Other cases are for the proposed solution with $\alpha\in \text{U}(\alpha_{min},\alpha_{max}$).
    }
    \label{fig:EER_errors}
    \vspace{-0.4cm}
\end{figure}

Anonymisation results in terms of ASV EER are illustrated in Figure~\ref{fig:EER_errors}.  Results are shown for an original setup without anonymisation, for the primary and secondary VoicePrivacy 2020 baselines (B1 and B2) and for B2 stochastic variants where the McAdams coefficient is drawn from different intervals in $\alpha \in \mbox{U}(\alpha_{min},\alpha_{max})$. All systems based on the McAdams coefficient have an LPC order of 20.  

It is clear that all systems increase the EER from the original 3.3\%.  Blue, left-most bars 
%in each group 
show EERs for ignorant privacy adversaries (o-a scenario) when test utterances are anonymised but enrolment utterances are not.  Results show that B1 out-performs B2 and all related systems by a substantial margin.  Yellow, middle bars show performance for semi-informed privacy adversaries (a-a scenario) when both enrolment and test utterances are anonymised (with the same, but differently configured algorithm).  Now, B2 and related systems are more competitive, e.g.\ 33.8\% for U(0.7,0.9) \emph{cf.}\ 32.3\% for B1.

\begin{figure}[!t]
    \centering
    \includegraphics[width=\columnwidth]{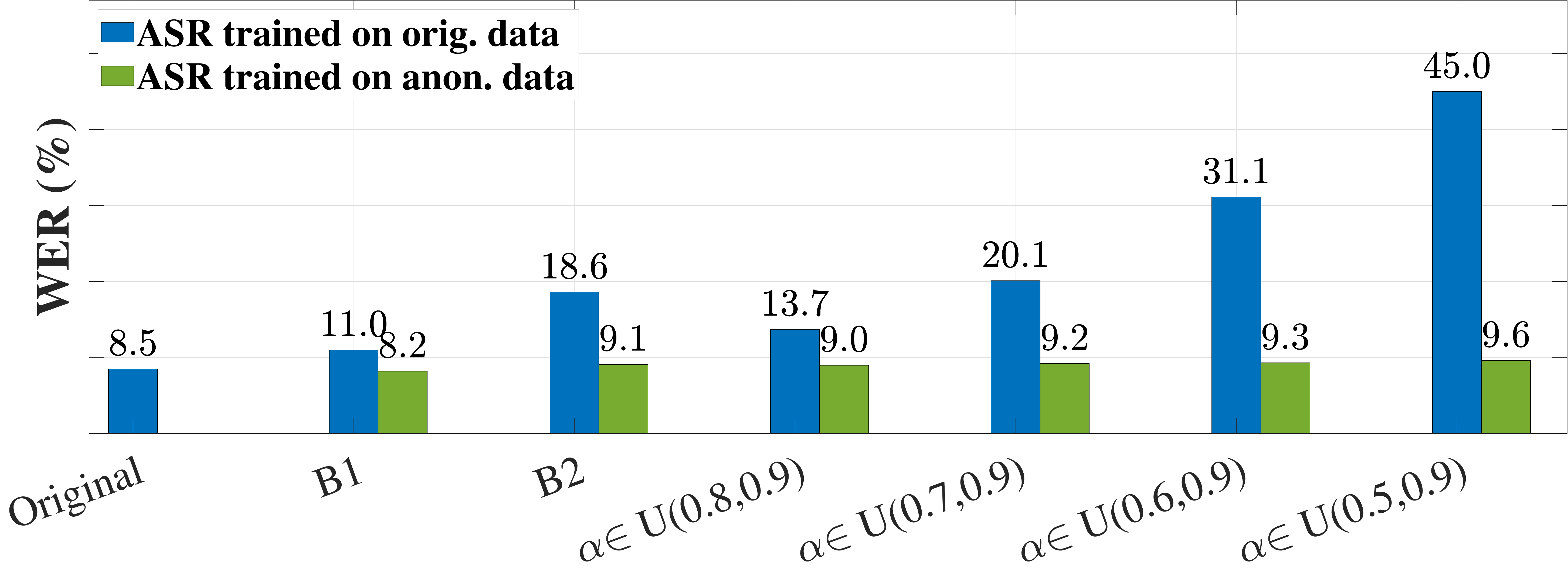}
    \caption{As for Fig.~\ref{fig:EER_errors} except for the ASR performance in terms of the word error rate (WER, \%).
    }
    \label{fig:WER_errors}
\end{figure}

While lower values of $\alpha_{min}$ give better anonymisation, improvements are accompanied by degradations to intelligibility.  ASR results for the same setups are illustrated in Figure~\ref{fig:WER_errors}.  An original, baseline WER of 8.45\% increases only to 10.95\% for B1, but to 13.7\% for U(0.8,0.9) and worse for other configurations.  Thus, while the McAdams approach to anonymisation can better protect privacy, it does so at the cost of reduced intelligibility.

The question then is what level of improvement to intelligibility can be delivered simply by retraining the ASR systems with anonymised speech and what would be the corresponding impacts to ASV.  Answers are provided by the green bars in Figures~\ref{fig:EER_errors} and~\ref{fig:WER_errors} for ASV and ASR systems respectively. For ASV, the McAdams approach to anonymisation performs best, e.g.\ 22.6\% for U(0.8,0.9) \emph{cf.}\ 13.7\% for B1.  Now, though, while better anonymisation is achieved for $\alpha_{min}<0.8$, the corresponding WER is much more competitive with the degradation for B1.  Whereas for the latter, the WER decreases from 10.95\% to 8.25\%, it decreases from 45\% to 9.6\% for U(0.5,0.9), while the EER is almost twice as high than for B1, e.g.\ up to 37.5\% for U(0.5,0.9).

Table~\ref{tab:zebra} shows ZEBRA results which reflect the evidence remaining to a privacy adversary after anonymisation.  They correspond to the semi-informed scenario where the ASV system is retrained on anonymised data, i.e.\ the green bars in Figures~\ref{fig:EER_errors}.   While the scales are inverted (less bits of disclosure infer better privacy), ZEBRA results shown in column 2 confirm much the same trend shown by EER results.  Noting that anonymisation affords different levels of privacy to different individuals, the ZEBRA framework also allows one to determine the \emph{worst case} level of privacy disclosure.  According to~\cite{nautsch2020privacy}, this is expressed according to categorical tags where a tag of `0' reflects no privacy disclosure, 'A' is the next best case and where tag `F' reflects the worst.  Identical categorical tags of `C' illustrated in column 3 for each system suggest that the differences in performance discussed above are not so great from the perspectives of a worst case scenario. 

\begin{table}[t]
\vspace{-0.1cm}
\centering
\caption{ZEBRA results for the semi-informed (a-a) scenario and where ASV systems are trained using anonymised data. 
}
\resizebox{0.65\columnwidth}{!}{%
\begin{tabular}{|c|c|c|}
\hline
\textbf{\begin{tabular}[c]{@{}c@{}}System\end{tabular}} &
  \textbf{\begin{tabular}[c]{@{}c@{}}Expected privacy\\disclosure (bits) \end{tabular}} &
  \textbf{\begin{tabular}[c]{@{}c@{}}Worst case\\ categorical tag\end{tabular}} \\ \hline
Original             & 0.647 & C \\ \hline
B1                   & 0.403 & C \\ \hline
B2                   & 0.598 & C \\ \hline
$\alpha\in(0.8,0.9)$ & 0.366 & C \\ \hline
$\alpha\in(0.7,0.9)$ & 0.261 & C \\ \hline
$\alpha\in(0.6,0.9)$ & 0.199 & C \\ \hline
$\alpha\in(0.5,0.9)$ & 0.155 & C \\ \hline
\end{tabular}%
}
\vspace{-0.5cm}

\label{tab:zebra}
\end{table}

%% file: sections/6_conclusions.tex
This paper shows that well-known signal processing techniques can provide efficient and effective solutions to anonymisation for speech signals.  The approach reported in this paper, based upon transformations to the spectral envelope using the McAdams coefficient, can increase by ten-fold the equal error rate of a baseline automatic speaker recognition system, with only modest degradations to intelligibility, even in a semi-informed privacy adversary scenario.  This is achieved with a randomised transformation which provides some level of protection from reversibility.

Despite this encouraging result, we must acknowledge that, until error rates increase to the equivalent of random performance, or a ZEBRA category of `0', we remain far from meeting the goal of \emph{true} anonymisation.  Database, gender and other biases mean that an `averaged' view of privacy tells an incomplete picture, especially when we know that even the stronger anonymisation solutions still leave some with relatively weaker protection.  Encouragingly, there is plenty of scope to extend this work.  Thus far we have considered only adjustments to the formant positions in terms of frequency.  Future work should explore more elaborate transformations to the spectral envelope which could be investigated through random pole perturbations. There is also scope to explore the resilience of the approach to brute-force attacks, and new approaches which combine the merits of our approach with those of the more sophisticated VoicePrivacy primary baseline based upon state-of-the-art x-vector embeddings and neural waveform techniques.

Last, all results reported in this paper are reproducible with open source code and scripts available online\footnote{\url{https://github.com/josepatino/Voice-Privacy-Challenge-2020/}}.

%% file: INTERSPEECH2021.bbl
% Generated by IEEEtran.bst, version: 1.13 (2008/09/30)
\begin{thebibliography}{10}
\providecommand{\url}[1]{#1}
\csname url@samestyle\endcsname
\providecommand{\newblock}{\relax}
\providecommand{\bibinfo}[2]{#2}
\providecommand{\BIBentrySTDinterwordspacing}{\spaceskip=0pt\relax}
\providecommand{\BIBentryALTinterwordstretchfactor}{4}
\providecommand{\BIBentryALTinterwordspacing}{\spaceskip=\fontdimen2\font plus
\BIBentryALTinterwordstretchfactor\fontdimen3\font minus
  \fontdimen4\font\relax}
\providecommand{\BIBforeignlanguage}[2]{{%
\expandafter\ifx\csname l@#1\endcsname\relax
\typeout{** WARNING: IEEEtran.bst: No hyphenation pattern has been}%
\typeout{** loaded for the language `#1'. Using the pattern for}%
\typeout{** the default language instead.}%
\else
\language=\csname l@#1\endcsname
\fi
#2}}
\providecommand{\BIBdecl}{\relax}
\BIBdecl

\bibitem{nautsch2019gdpr}
A.~Nautsch, C.~Jasserand, E.~Kindt, M.~Todisco, I.~Trancoso, and N.~Evans,
  ``{The GDPR \& speech data: Reflections of legal and technology communities,
  first steps towards a common understanding},'' in \emph{Proc. INTERSPEECH},
  2019.

\bibitem{nautsch2019preserving}
A.~Nautsch, A.~Jim{\'e}nez, A.~Treiber, J.~Kolberg, C.~Jasserand, E.~Kindt,
  H.~Delgado, M.~Todisco, M.~A. Hmani, A.~Mtibaa \emph{et~al.}, ``Preserving
  privacy in speaker and speech characterisation,'' \emph{Computer Speech \&
  Language}, vol.~58, pp. 441--480, 2019.

\bibitem{tomashenko2020introducing}
N.~Tomashenko, B.~M.~L. Srivastava, X.~Wang, E.~Vincent, A.~Nautsch,
  J.~Yamagishi, N.~Evans, J.~Patino, J.-F. Bonastre, P.-G. No{\'e}
  \emph{et~al.}, ``{Introducing the VoicePrivacy initiative},'' in \emph{Proc.
  INTERSPEECH}, 2020.

\bibitem{tomashenko2020voiceprivacy}
------, ``{The VoicePrivacy 2020 Challenge evaluation plan},'' 2020.

\bibitem{noe2020speech}
P.-G. No{\'e}, J.-F. Bonastre, D.~Matrouf, N.~Tomashenko, A.~Nautsch, and
  N.~Evans, ``{Speech Pseudonymisation Assessment Using Voice Similarity
  Matrices},'' in \emph{Proc. INTERSPEECH}, 2020.

\bibitem{fangspeaker}
F.~Fang, X.~Wang, J.~Yamagishi, I.~Echizen, M.~Todisco, N.~Evans, and J.-F.
  Bonastre, ``{Speaker Anonymization Using X-vector and Neural Waveform
  Models},'' in \emph{Proc. 10th ISCA Speech Synthesis Workshop}, 2018, pp.
  155--160.

\bibitem{mcadams1984spectral}
S.~McAdams, ``Spectral fusion, spectral parsing and the formation of the
  auditory image,'' \emph{Ph. D. Thesis, Stanford}, 1984.

\bibitem{patino2020mcadams}
J.~Patino, M.~Todisco, A.~Nautsch, and N.~Evans, ``{Speaker anonymisation using
  the McAdams coefficient},'' Eurecom, Tech. Rep. RR-20-343. 2020 [Online].
  Available: http://www.eurecom.fr/publication/6190, Tech. Rep., 2020.

\bibitem{gupta2020design}
P.~Gupta, G.~P. Prajapati, S.~Singh, M.~R. Kamble, and H.~A. Patil, ``Design of
  voice privacy system using linear prediction,'' in \emph{Proc. APSIPA}.\hskip
  1em plus 0.5em minus 0.4em\relax IEEE, 2020, pp. 543--549.

\bibitem{kai2021slt}
H.~Kai, S.~Takamichi, S.~Shiota, and H.~Kiya, ``Lightweight voice anonymization
  based on data-driven optimization of cascaded voice modification modules,''
  in \emph{Proc. IEEE SLT}, 2021.

\bibitem{computermusic}
C.~Dodge and T.~A. Jerse, \emph{Computer Music: Synthesis, Composition and
  Performance}, 2nd~ed.\hskip 1em plus 0.5em minus 0.4em\relax Macmillan
  Library Reference, 1997.

\bibitem{jin2009speaker}
Q.~Jin, A.~R. Toth, T.~Schultz, and A.~W. Black, ``Speaker de-identification
  via voice transformation,'' in \emph{Proc. ASRU}.\hskip 1em plus 0.5em minus
  0.4em\relax IEEE, 2009, pp. 529--533.

\bibitem{pobar2014online}
M.~Pobar and I.~Ip{\v{s}}i{\'c}, ``Online speaker de-identification using voice
  transformation,'' in \emph{2014 37th International convention on information
  and communication technology, electronics and microelectronics
  (mipro)}.\hskip 1em plus 0.5em minus 0.4em\relax IEEE, 2014, pp. 1264--1267.

\bibitem{bahmaninezhad2018convolutional}
F.~Bahmaninezhad, C.~Zhang, and J.~H. Hansen, ``Convolutional neural network
  based speaker de-identification.'' in \emph{Proc. Odyssey}, 2018, pp.
  255--260.

\bibitem{qian2017voicemask}
J.~Qian, H.~Du, J.~Hou, L.~Chen, T.~Jung, X.-Y. Li, Y.~Wang, and Y.~Deng,
  ``Voicemask: Anonymize and sanitize voice input on mobile devices,''
  \emph{arXiv preprint arXiv:1711.11460}, 2017.

\bibitem{magarinos2017reversible}
C.~Magari{\~n}os, P.~Lopez-Otero, L.~Docio-Fernandez, E.~Rodriguez-Banga,
  D.~Erro, and C.~Garcia-Mateo, ``Reversible speaker de-identification using
  pre-trained transformation functions,'' \emph{Computer Speech \& Language},
  vol.~46, pp. 36--52, 2017.

\bibitem{srivastava2019privacy}
B.~M.~L. Srivastava, A.~Bellet, M.~Tommasi, and E.~Vincent,
  ``{Privacy-preserving adversarial representation learning in ASR: Reality or
  illusion?}'' \emph{arXiv preprint arXiv:1911.04913}, 2019.

\bibitem{cohen1995vocal}
J.~Cohen, T.~Kamm, and A.~G. Andreou, ``{Vocal tract normalization in speech
  recognition: Compensating for systematic speaker variability},'' \emph{The
  Journal of the Acoustical Society of America}, vol.~97, no.~5, pp.
  3246--3247, 1995.

\bibitem{lee1998frequency}
L.~Lee and R.~Rose, ``A frequency warping approach to speaker normalization,''
  \emph{IEEE Transactions on speech and audio processing}, vol.~6, no.~1, pp.
  49--60, 1998.

\bibitem{qian2018hidebehind}
J.~Qian, H.~Du, J.~Hou, L.~Chen, T.~Jung, and X.-Y. Li, ``{Hidebehind: Enjoy
  Voice Input with Voiceprint Unclonability and Anonymity},'' in \emph{Proc.
  ACM Conference on Embedded Networked Sensor Systems}, 2018, pp. 82--94.

\bibitem{srivastava2020evaluating}
B.~M.~L. Srivastava, N.~Vauquier, M.~Sahidullah, A.~Bellet, M.~Tommasi, and
  E.~Vincent, ``Evaluating voice conversion-based privacy protection against
  informed attackers,'' in \emph{Proc. ICASSP}.\hskip 1em plus 0.5em minus
  0.4em\relax IEEE, 2020, pp. 2802--2806.

\bibitem{srivastava2020design}
B.~M.~L. Srivastava, N.~Tomashenko, X.~Wang, E.~Vincent, J.~Yamagishi,
  M.~Maouche, A.~Bellet, and M.~Tommasi, ``Design choices for x-vector based
  speaker anonymization,'' in \emph{Proc. INTERSPEECH}, 2020.

\bibitem{panayotov2015librispeech}
V.~Panayotov, G.~Chen, D.~Povey, and S.~Khudanpur, ``{Librispeech: an ASR
  corpus based on public domain audio books},'' in \emph{Proc. ICASSP}.\hskip
  1em plus 0.5em minus 0.4em\relax IEEE, 2015, pp. 5206--5210.

\bibitem{veaux2016vctk}
C.~Veaux, J.~Yamagishi, K.~MacDonald \emph{et~al.}, ``{CSTR VCTK corpus:
  English multi-speaker corpus for CSTR Voice Cloning Toolkit},'' 2016.

\bibitem{nautsch2020privacy}
A.~Nautsch, J.~Patino, N.~Tomashenko, J.~Yamagishi, P.-G. Noe, J.-F. Bonastre,
  M.~Todisco, and N.~Evans, ``{The Privacy ZEBRA: Zero Evidence Biometric
  Recognition Assessment},'' in \emph{Proc. INTERSPEECH}, 2020.

\end{thebibliography}
